\documentclass[a4paper,11pt]{article}
\usepackage{aaskaiid}
 \usepackage{orcidlink}

\title{Exploring the physics behind the observed magnetic filaments in large scale radio galaxies }
\ShortTitle{Physics of magnetic filaments}

\author[1]{Christian M. Fromm\orcidlink{0000-0002-1827-1656}}
\ShortName{Fromm et al.} 
\author[1]{Matthias Kadler\orcidlink{0000-0001-5606-6154}}
\author[1]{Karl Mannheim\orcidlink{0000-0002-2950-6641}}

\affiliation[1]{Institut f\"ur Theoretische Physik und Astrophysik, Universit\"at W\"urzburg, Emil-Fischer-Str. 31, D-97074 W\"urzburg, Germany}
\emailAdd{christian.fromm@uni-wuerzburg.de}
\emailAdd{matthias.kadler@uni-wuerzburg.de}
\emailAdd{karl.mannheim@uni-wuerzburg.de}

\abstract{Recent low-frequency MeerKAT observations of radio galaxies have revealed an unexpected population of thin, highly collimated synchrotron threads, whose numbers continue to grow with increasing survey depth and sensitivity. These intricate structures display a remarkable diversity of morphologies—appearing as narrow filaments linking jets and lobes, as well as ring- or ribbon-like features embedded within the jets and radio lobes. Despite their ubiquity, the physical origin and stability of these collimated synchrotron threads remain poorly understood. Proposed mechanisms include shock compression and interactions with the magneto-ionic intracluster medium, magnetic flux tube formation, and reconnection-driven magnetic filaments.

In this work, we investigate the formation and evolution of such magnetic filaments using three-dimensional, two-temperature general relativistic magnetohydrodynamic (GRMHD) simulations of realistically launched jets from supermassive black holes. From these simulations, we compute synthetic synchrotron emission maps and polarisation signatures, allowing us to predict the observable characteristics—including morphology, brightness profiles, and polarisation patterns. Finally, we assess the detectability and diagnostic potential of these signatures with the Square Kilometre Array Observatory (SKAO), outlining how upcoming SKA observations can distinguish between competing physical models and illuminate the magnetic origin of the collimated synchrotron threads revealed by MeerKAT observations.}

\begin{document}
\maketitle
\
\section{Introduction}
Deep radio observations with the MeerKAT array have revealed a new population of extremely narrow, elongated synchrotron features in radio galaxies and cluster cores, indicating an unexpected degree of magnetic order and high polarisation fractions of up to 20$\%$. These \emph{collimated synchrotron threads} (CSTs) are characterised by lengths of tens to hundreds of kiloparsecs, widths typically below a kiloparsec, and steep synchrotron spectra ($\alpha \simeq 1.5$--2.0). The most prominent example is found in ESO~137$-$006, located at the centre of the Norma cluster, where a $\sim$80\,kpc-long and $\sim$1\,kpc-wide thread connects the two bent radio lobes of the source \citep{Ramatsoku2020}. Similar linear filaments have been detected in IC~4296, together with ribbon- and ring-like features in the radio lobes \citep{Condon2021}, while fainter analogues have been reported in the Ophiuchus cluster \citep{Botteon2025} and in A3562 \citep{Venturi2022}. Highly polarised threads, apparently extending perpendicular to the jet axis and streaming into the intracluster medium (ICM), have also been identified in 3C\,40B through the MeerKAT Galaxy Cluster Legacy Survey and the LOFAR Two-Metre Sky Survey \citep{Rudnick2022}. Together, these discoveries suggest that such structures may be widespread but have remained undetected until now due to the sensitivity and sub-arcminute resolution required to resolve them.

The morphology of CSTs is remarkable: their aspect ratios often exceed $50{:}1$, implying long, well-collimated channels of magnetised plasma embedded in the ICM. In ESO~137$-$006 and IC~4296, the threads lie outside the main radio lobes, apparently bridging them across projected distances of $\sim$50–80\,kpc while maintaining coherent brightness and a smoothly steepening spectral index along their lengths. In IC~4296, the emission exhibits strong linear polarisation with magnetic field vectors largely aligned with the thread axis, further implying a high degree of field ordering over tens of kiloparsecs.

The physical origin of these filaments remains uncertain, and several mechanisms have been proposed to explain their formation and longevity. One possibility is that they arise from shear-driven amplification and ordering of magnetic fields along jet or cocoon boundaries, where velocity gradients excite Kelvin–Helmholtz (KH) or current-driven instabilities that stretch and align magnetic field lines, forming narrow synchrotron-emitting channels \citep[see for a review][]{Perucho2019}. A complementary mechanism invokes magnetised back-flows from the radio lobes: as the AGN jets inflate lobes and drive cocoon expansion, an internal back-flow of plasma can drag magnetic flux downstream into the ICM, creating narrow, collimated threads of magnetised plasma trailing the lobes \citep[see, e.g.,][]{Upreti2024}. Yet another line of thought appeals to the dynamics of the central engine itself: when the accretion flow onto a super-massive black hole accumulates sufficient poloidal magnetic flux it enters a magnetically arrested disk (MAD) state \citep{Tchekhovskoy2011}, drastically boosting jet power and organising the magnetic field geometry; episodic flux eruptions or field reconfigurations in the MAD regime may launch magnetised, stable channels that survive long distances and could seed CSTs \citep{Porth2021}. In addition, threads may trace weak shocks or compression fronts produced as expanding AGN cocoons interact with the ambient ICM, leading to magnetic field amplification and local re-acceleration of relativistic electrons via diffusive shock acceleration  \citep{Alam2025}. In some cases, the threads might correspond to magnetised plasma filaments generated by turbulence, shear flows or magnetic reconnection in the wake of radio jets \citep{Rudnick2022}.  

In this work, we investigate whether magnetised back flow and/or episodic ejections from MAD systems can account for the observed CSTs. We employ global large-scale GRMHD simulations of spinning black holes coupled with kinetic prescriptions for particle acceleration, to model the evolution, confinement, and synchrotron emissivity of magnetised flux tubes as they propagate into a strafified ambient medium environment. Synthetic synchrotron maps derived from these simulations are compared with high-resolution MeerKAT and VLBI data of edge-brightened jets and filaments. This approach aims to bridge the gap between black-hole-scale magnetic dynamics and the kiloparsec-scale radio morphology of active galaxies, providing a unified physical framework for the formation and persistence of collimated synchrotron threads.

\section{Numerical Setup}

\subsection{GRMHD Simulations}

We perform three-dimensional two-temperature general relativistic magnetohydrodynamic (GRMHD) simulations using the \texttt{KHARMA} code in modified Kerr–Schild coordinates \citep[see][for details]{Prather2024}. Throughout this work, we adopt code units for time and length scales: the gravitational light-crossing time is set as $t_{\mathrm{g}} = G M_\mathrm{BH}/c^3 \equiv 1\,{M}$, and the gravitational radius as $r_\mathrm{g} = G M_\mathrm{BH}/c^2 \equiv 1\,{M}$, where $G$ is the gravitational constant, $c$ the speed of light in vacuum, $M_\mathrm{BH}$ the black hole mass, and ${M}$ the code mass unit. The computational grid is resolved with $(N_r, N_\theta, N_\phi) = (384, 192, 128)$ in the respective radial, polar, and azimuthal directions. The radial domain spans $r \in [1.1659, 10^4]\,{M}$, while the full angular domains are $\theta \in [0, \pi]$ and $\phi \in [0, 2\pi]$. The simulation is initialized with a Fishbone–Moncrief torus \citep{FM1976} in hydrostatic equilibrium, characterized by an inner edge at $r_\mathrm{in}=20\,{M}$ and a pressure maximum at $r_\mathrm{c}=40\,{M}$. The torus is threaded by a weak poloidal magnetic field, specified via the vector potential 
\begin{equation}\label{eq:vectorpotential}
A_\phi = \max\left[\frac{\rho}{\rho_\mathrm{max}}\left(\frac{r}{r_\mathrm{in}}\right)^3 \sin^3\theta \exp\left(-\frac{r}{400}\right) - 0.01,\, 0\right],
\end{equation}
and normalized to a plasma beta of $\beta = 100$. This form of the vector potential and accretion disk will lead to Magnetically Arrest Disk scenario (MAD) \citep{Tchekhovskoy2011}. We use a black hole with spin of  $a_* = 0.94$ and the fluid in the simulation follows an ideal gas equation of state with adiabatic index $\Gamma = 5/3$ appropriate for relativistic plasma under conditions relevant to low luminosity black
hole accretion and assuming that ions dominate the energy and pressure budget \citep[see for details][]{Gammie2025}. During the run time of the simulations we include particle heating due to magnetic reconnection and turbulence \citep[see, e.g.,][for details]{Mizuno2021}. Following \citet{Rohoza2024,Lalakos2024}, we rescaled the Bondi radius to $R_\mathrm{B} = G M_\mathrm{BH}/c_\infty^2 = 1000\,{M}$ to capture the influence of the ambient medium and to enable the large-scale evolution of the jet, including self-consistent jet launching from the accretion disk within our numerical grid. Inside the Bondi radii, we  apply our floor model while outside we fix the ambient density to $\rho=10^{-5}$.

\subsection{GRRT calculations}
\subsection{General Relativistic Radiative Transfer}\label{sec:GRRT}

To connect our 3D GRMHD simulations with observable signatures, we perform polarized synchrotron radiative transfer in post-processing using the semi-analytic code \texttt{IPOLE} \citep{Monika2018}. The electron temperature is computed from the electron entropy, evolved during the GRMHD simulations $k_{\rm{e}}$ according to:
\begin{equation}
    \Theta_{\rm e} =k_{\rm e}\rho^{\hat{\gamma_{\rm e}}-1}\frac{m_{\rm p}}{m_{\rm e}},
\end{equation}
where $\hat{\gamma_{\rm e}}=4/3$ is the adiabatic index of the electrons and $m_{\rm p}$ is the proton mass. During the radiative transfer we assume that the electron distribution function (eDF) is thermal (Maxwell-Jüttner) in the dense disk regions:
\begin{equation}
    \frac{1}{n_e}\frac{dn_e}{d\gamma_\mathrm{e}} = \frac{\gamma_\mathrm{e}^2 \sqrt{1-1/\gamma_\mathrm{e}^2}}{\Theta_e K_2(1/\Theta_e)},
\end{equation}
with $\Theta_e=k_\mathrm{B}T_e/m_e c^2$ and $K_2$ the modified Bessel function. While in the jet, where non-thermal acceleration is expected due to magnetic reconnection and turbulence, we adopt a $\kappa$-distribution \citep{Pandya2016}:
\begin{equation}
    \frac{1}{n_e}\frac{dn_e}{d\gamma_\mathrm{e}} = N \gamma_\mathrm{e} \sqrt{\gamma_\mathrm{e}^2-1} \left(1+\frac{\gamma_\mathrm{e}-1}{\kappa w}\right)^{-(\kappa+1)}.
\end{equation}
Here, the parameters $\kappa$ and $w$ are determined from PIC-motivated subgrid models that relate the particle distribution to local plasma properties, including magnetization $\sigma$ and plasma-$\beta$ \citep{Fromm2022,Meringolo2023}:
\begin{align}
    \kappa &= 2.8 + 0.2\sigma^{-1/2} + 1.6 \sigma^{-6/10} \tanh(2.25 \sigma^{1/3}\beta),\\
    w &= \frac{\kappa-3}{\kappa}\Theta_e + \frac{\varepsilon}{2}[1+\tanh(r-r_\mathrm{inj})]\frac{\kappa-3}{6\kappa}\frac{m_\mathrm{p}}{m_\mathrm{e}}\sigma,\\
    \varepsilon &= 1.0 - 0.23 \sigma^{-1/2} + 0.5 \sigma^{1/10} \tanh(-10.18 \beta \sigma^{1/10}),
\end{align}
with the non-thermal injection radius set to $r_\mathrm{inj}=10\,{M}$. The $\kappa$-distribution is applied only for $3.1<\kappa<8$, smoothly converging to the Maxwell-Jüttner distribution outside this range.

\subsubsection{Anisotropic Synchrotron Emission}

Electrons in weakly collisional or collisionless plasmas develop pitch-angle anisotropies due to synchrotron cooling and kinetic instabilities \citep{Zhdankin2023,Comisso2024}. In particular, the synchrotron firehose instability can redistribute particle momenta, resulting in a preferential alignment along the magnetic field. Observational and theoretical studies suggest that such anisotropies are relevant in relativistic jets \citep{Tsunetoe2025}. To model this effect, we expand the isotropic $\kappa$-distribution emissivities $j_{a,\,\rm iso}$ and absorption coefficients $\alpha_{a,\,\rm iso}$ around the pitch angle $\theta_B$ between the magnetic field and photon propagation:
\begin{align}
    j_a &= j_{a,\,\rm iso}\, \phi(\theta_B), &
    j_V &= j_{V,\,\rm iso}\, \phi(\theta_B) \left( 1+ \frac{g(\theta_B)}{p+2}\right),\\
    \alpha_a &= \alpha_{a,\,\rm iso}\, \phi(\theta_B), &
    \alpha_V &= \alpha_{V,\,\rm iso}\, \phi(\theta_B) \left( 1+ \frac{g(\theta_B)}{p+2}\right),
\end{align}
where
\begin{align}
    \phi(\theta_B) &= P(p,\eta)^{-1} [1+(\eta-1)\cos^2\theta_B]^{-p/2},\\
    g(\theta_B) &= \frac{p(\eta-1)\sin^2\theta_B}{1+(\eta-1)\cos^2\theta_B},\\
    P(p,\eta) &= \int_0^1 d\mu [1+(\eta-1)\mu^2]^{-p/2},
\end{align}
and $p=\kappa-1$. This approach allows us to capture the combined effects of non-thermal particle acceleration and relativistic beaming along magnetic field lines, producing a realistic representation of anisotropic jet emission.  Overall, this framework enables a self-consistent calculation of polarized synchrotron emission from the accretion disk and jet, connecting GRMHD dynamics with observables at VLBI scales.

\section{Results}
\subsection{GRMHD}
\begin{figure}[h]
    \centering
	\includegraphics[width=0.99\columnwidth]{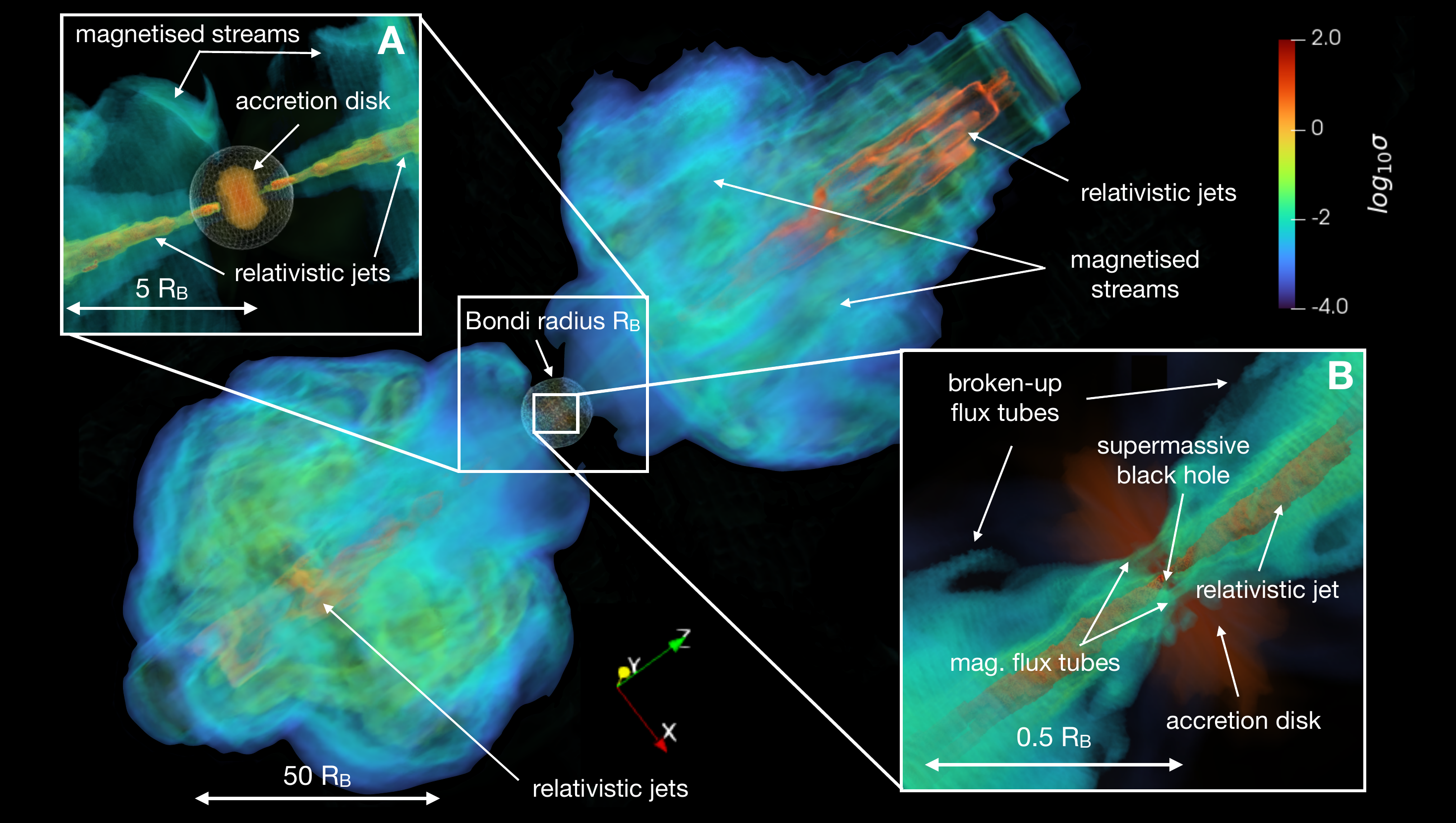}
    \caption{3D Volume rendering of the magnetisation from a GRMHD snapshot at $t = 40\,{\rm kM}$}
    \label{fig:3Drender}
\end{figure}

In Fig.~\ref{fig:3Drender}, we present a three-dimensional volume rendering of the magnetisation parameter, $\sigma = B^2 / \rho$, from a general-relativistic magnetohydrodynamic (GRMHD) simulation snapshot at $t = 40\,{\rm kM}$. During the early stages of the simulation ($t < 10\,{\rm kM}$), the black hole accretes both mass and magnetic flux from the surrounding accretion disk, leading to the build-up of a dynamically important poloidal magnetic field near the event horizon. As a result, a relativistic jet is launched via the Blandford--Znajek mechanism \citep{Blandford1977}, extracting rotational energy from the spinning black hole.  

Initially, the jet propagates within a low-density environment corresponding to the numerical floor model inside the Bondi radius ($r < R_{\rm B}$). In this region, the outflow expands freely and develops a parabolic geometry, consistent with magnetically accelerated jet collimation profiles \citep[e.g.][]{Tchekhovskoy2011, Porth2021,Fromm2022}. Once the jet exits the low-density zone ($r > R_{\rm B}$), it interacts with a denser ambient medium, giving rise to a complex jet–cocoon structure. This structure consists of a dilute, highly magnetised relativistic jet spine surrounded by a denser, weakly magnetised sheath propagating in a dense weakly magnetised cocoon formed by shocked ambient gas and entrained material (see light green features in Fig.~\ref{fig:3Drender}). At the jet–sheath interface, velocity shear and density gradients trigger Kelvin–Helmholtz and Rayleigh–Taylor instabilities, driving turbulence and mixing within the cocoon \citep[e.g.][]{Perucho2019}. We find that the jet spine remains largely sub-Alfvénic near the black hole $(v \leq v_A)$, while the sheath and wind become super-Alfvénic, supporting the development of Kelvin–Helmholtz instabilities at the jet–sheath interface, consistent with the linear analyses of \cite{Bodo2013,Bodo2019}.

\begin{figure}[h]
    \centering
	\includegraphics[width=0.99\columnwidth]{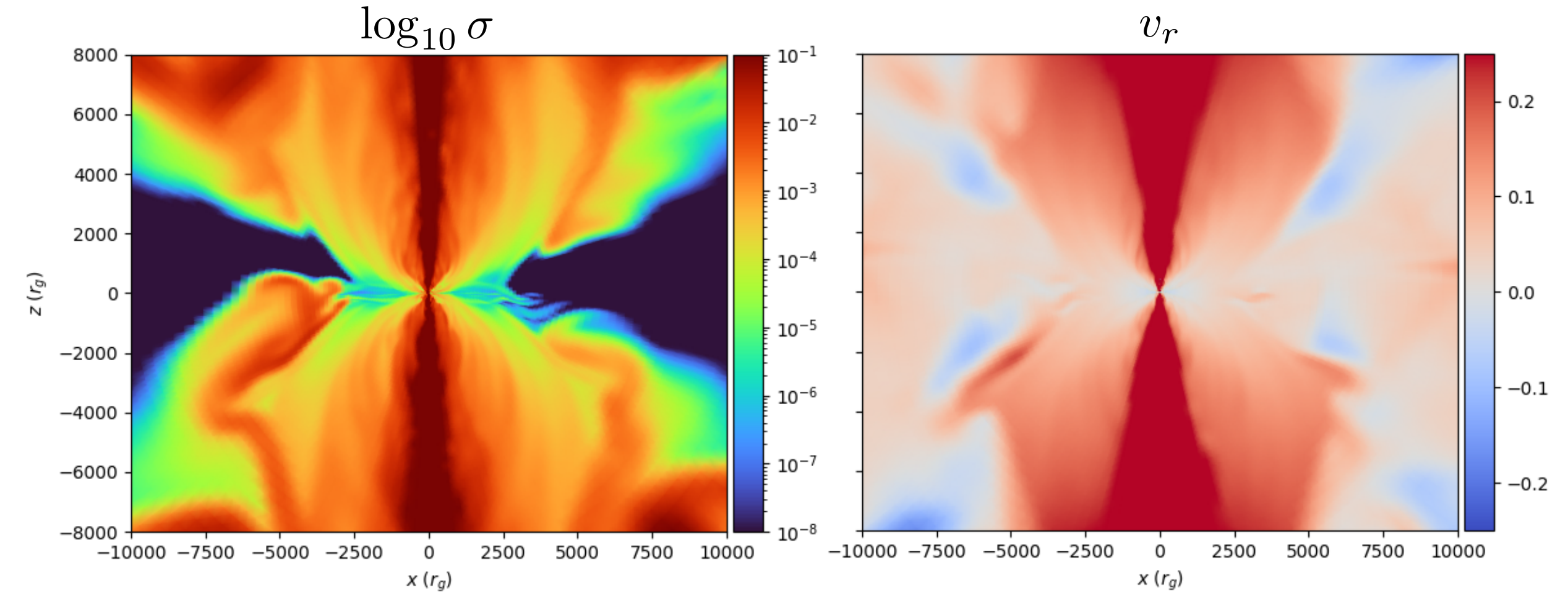}
    \caption{Distribution of magnetisation, $\sigma$, (left) and radial velocity, $v_r$, (right) in the meridional plane}
    \label{fig:sigmavr}
\end{figure}

These interactions lead to the formation of magnetised plasma streams and vortical motions within the cocoon. In addition to the forward-propagating jet, the simulations reveal the presence of coherent, magnetised back-flows, streams of plasma moving in the reverse direction toward the black hole (see inset A in Fig.~\ref{fig:3Drender}) . These back-flows arise naturally from the pressure imbalance and turbulent recirculation within the cocoon, carrying magnetic flux and energy back toward the central engine potentially seeding the formation of collimated, magnetised filaments in the outer jet regions. In Fig.~\ref{fig:sigmavr} we show a zoom of the inner $\sim 10^{4}\,r_{\rm g}$ highlighting the magnetisation, $\sigma$, (left) and the radial component of the four-velocity, $v_r$, (right). 

The left panel reveals the strongly magnetised jet spine (dark red), surrounded by a network of collimated magnetic filaments with progressively lower magnetisation (light red to orange), and finally by weakly or un-magnetised ambient material entrained through Kelvin--Helmholtz--driven mixing (green to blue). When comparing these structures with the radial velocity map, we find that the outermost magnetic filaments with intermediate magnetisation ($10^{-3} \lesssim \sigma \lesssim 10^{-2}$) exhibit a coherent inward radial motion (blue colours). These filaments are therefore likely to be re-accreted onto the black hole, potentially increasing the magnetic flux threading the horizon and thereby strengthening the jet. Notably, the morphology and coherence of these simulated filaments could resemble the collimated synchrotron threads revealed by recent \textit{MeerKAT} observations of radio galaxies, suggesting a common physical origin.  

During the course of the simulations, the black hole accretes a substantial amount of magnetic flux, leading to the build-up of a magnetic pressure barrier that can balance or even exceed the ram pressure of the inflowing matter. This process can displace parts of the accretion disk, resulting in the formation of magnetic flux tubes \citep{Porth2021}. These flux tubes are characterised by a strong vertical component of magnetic field and low plasma density, and they are advected within the disk while being pushed to larger distances from the black hole. While some flux tubes remain well-collimated over large distances ($r \gtrsim 1000\,r_{\rm g}$), others fragment during their outward propagation due to interactions with the surrounding disk and turbulent corona (see inset B in Fig.~\ref{fig:3Drender}). Such coherent, magnetised structures provide a natural channel for the transport of magnetic flux and may serve as seeds for large-scale, collimated filaments observed in extragalactic jets.

\subsection{GRRT}
\begin{figure}[h]
    \centering
	\includegraphics[width=1\columnwidth]{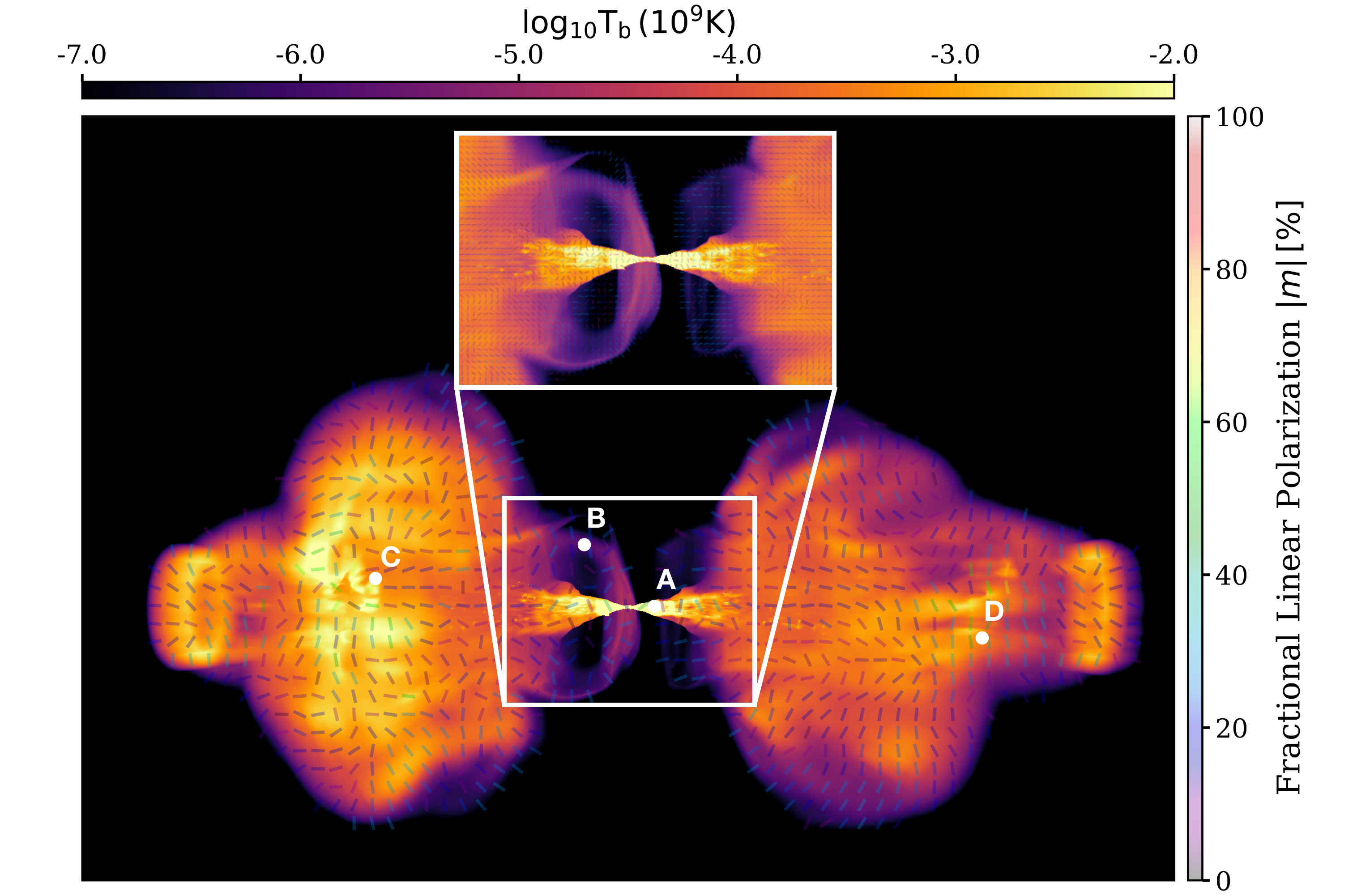}
    \caption{1.4 GHz polarised emission computed from our large-scale GRMHD simulations}
    \label{fig:emission}
\end{figure}

To compare our GRMHD model with recent MeerKAT observations and future SKAO data, we computed the corresponding synchrotron emission using general-relativistic radiative transfer (GRRT) calculations. Figure~\ref{fig:emission} shows the simulated 1.4\,GHz radio images derived from the GRMHD snapshot. We adopt a black hole mass of $\sim 10^{10}\,M_\odot$ \citep{Alam2025} and a viewing angle of $85^{\circ}$ and an anisotropy of $\eta=0.3$. The jet emission is modelled using a hybrid particle distribution: regions with relativistic outflows, identified by the Bernoulli criterion $-h u_t > 1.02$, are populated with non-thermal electrons, while thermal distributions are assumed elsewhere.  
The resulting total-intensity and polarisation maps reveal a complex emission morphology, including filamentary streams, arcs, and ring-like features that mirror the dynamics seen in the underlying GRMHD simulations. The overlaid electric vector position angles (EVPAs) trace these structures closely, indicating the geometry of the ordered magnetic field within the jet and cocoon. Notably, a magnetised back-flowing stream is clearly visible in both total intensity and polarisation: its EVPAs are aligned along its length, confirming that the emission originates from a coherent magnetic structure. A zoom into this region shows that the back-flow consists of multiple thin, collimated streams, some of which fragment into open, thread-like features mimicking those observed in ESO~137$-$006 and IC~4296.  

The simulated jet exhibits a pronounced spine–sheath structure, with the inner spine remaining highly magnetised and the sheath showing signs of interaction with the ambient medium. Small-scale wiggles and bends are apparent along the jet axis, likely induced by Kelvin–Helmholtz instabilities at the jet–ambient medium interface and possibly enhanced by current-driven instabilities in the magnetised spine. Together, these results demonstrate that GRMHD-driven jets naturally produce polarised, thread-like emission structures that are consistent with observations of radio galaxies.

\begin{figure}[h]
    \centering
	\includegraphics[width=1\columnwidth]{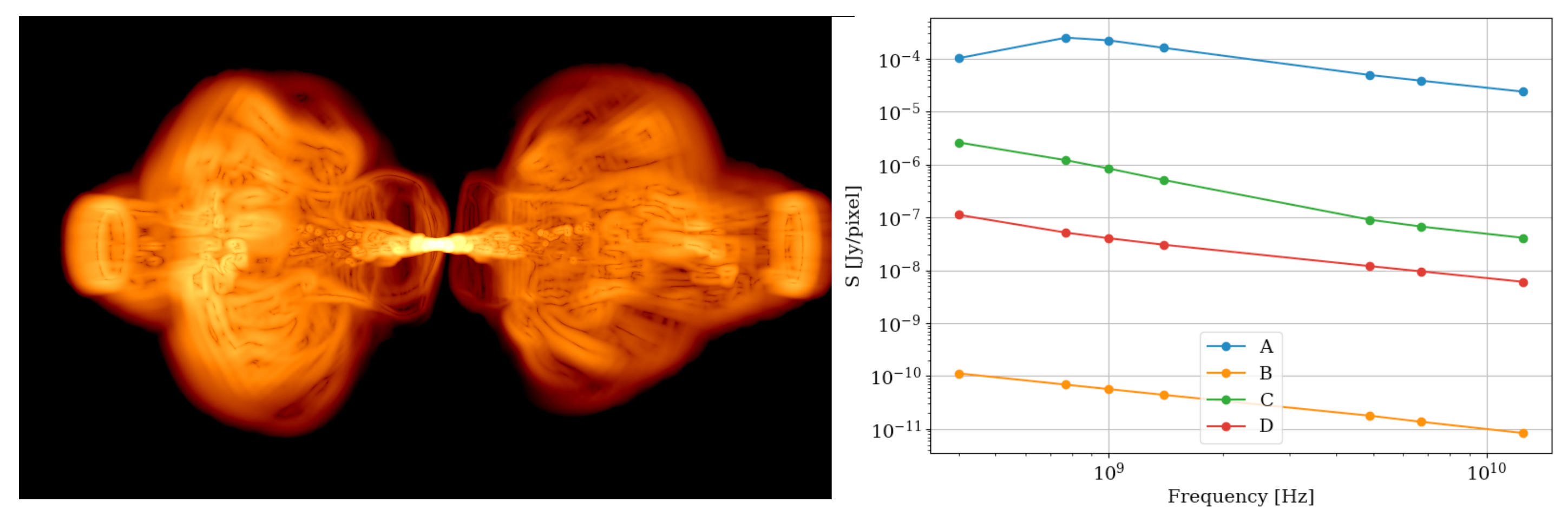}
    \caption{Sobel-edge-enhancement filter applied to GRRT image (left) and spectrum across SKA-MID frequencies at selected positions (see white points and labels in Fig. \ref{fig:emission}) on the right.}
    \label{fig:sobelspec}
\end{figure}

To enhance the visibility of the filamentary structure seen in our GRRT image we apply a Sobel edge-enhancement filter in Fig.\ref{fig:sobelspec} \citep[see also][]{Ramatsoku2020}. The result of the convolution highlights a rich network of fine, elongated filamentary structures embedded within the broader synchrotron-emitting jet lobes. These filaments trace regions where the intensity gradient is locally maximised, corresponding to sharp transitions in emissivity that arise at sites of strong shear, magnetic compression, and possible sites of magnetic reconnection within the turbulent jet plasma. The resulting pattern closely resembles the complex system of collimated synchrotron threads revealed by recent \textit{MeerKAT} observations of nearby radio galaxies \citep[e.g.][]{Ramatsoku2020}, where narrow, jet-aligned features extend far from the central engine and maintain coherence over several kiloparsecs. In particular, the alternating bright and faint ridges visible in our filtered GRRT image echo the morphology of the ``threads'' identified in the MeerKAT maps, suggesting that similar magnetohydrodynamic processes potentially associated with Kelvin–Helmholtz–driven shear layers, intermittent flux tubes, or dynamically advected magnetic filaments operate in both the simulations and the observed systems. The resemblance between the synthetic and observed structures provides further support for the interpretation that large-scale, ordered magnetic filaments are a natural outcome of relativistic jet evolution in a stratified ambient medium.

In right panel of Figure \ref{fig:sobelspec} displays the spectrum across the SKA-MID AA* bands for four key locations (A–D) within the jet, cocoon structure and magnetic back-flow (see Fig. \ref{fig:emission}). The spectra extracted across the SKA-MID AA* frequency bands  show clear differences in spectral behavior linked to the local plasma conditions. Location A, situated at the jet spine near the core, exhibits a relatively flat spectral index, consistent with optically thick synchrotron emission from freshly accelerated relativistic electrons in the highly magnetized jet base. In contrast, location B positioned just outside the jet boundary in a back-flowing magnetic filament shows a steep spectrum indicative lower magnetic field strength and density. Locations C and D, embedded within the filamentary structures of the cocoon, display intermediate spectral indices that suggest ongoing particle acceleration, potentially linked to the turbulent magnetic reconnection heating included in GRMHD and GRRT model (see Eqns 2 and 5-7). The broadband coverage of the AA* configuration will enable us to robustly quantify these spectral slopes, thus providing a powerful diagnostic for disentangling the physical processes shaping the complex emission structures in the lobes, as well as the magnetic filaments and the collimated synchrotron threads obtained with future SKA observations.

\section{Discussion \& Conclusion}
We have performed global, three-dimensional, two-temperature GRMHD simulations to investigate the physical origin of collimated synchrotron threads (CSTs) in radio galaxies. Our study focused on the evolution of magnetised plasma streams within the jet cocoon and back-flow regions, as well as on the formation of magnetic flux tubes expelled from the accretion disk during magnetically arrested disk (MAD) episodes. The simulations reveal multiple narrow, highly magnetised streams developing in the jet lobes and cocoon, including coherent back-flowing structures that transport magnetic flux and energy toward the central black hole. Notice, that magnetised large-scale jet injection simulations naturally produce rib-like and tethered structures through the growth of current-driven kink instabilities, whose nonlinear evolution dissipates magnetic energy and generates filamentary features resembling the observed MeerKAT morphology \citep{Upreti2024}.

Radiative transfer calculations based on our GRMHD data reproduce thin, polarised synchrotron filaments both within the lobes and in the magnetised back-flow, closely resembling the morphology and polarisation properties of the threads observed in ESO~137$-$006 and IC~4296. These results suggest that CST-like structures may naturally arise from the interplay of jet–cocoon dynamics, back-flow turbulence, and magnetic flux ejection during MAD events.  

Future work on the simulations will explore more realistic cluster environments, including a clumpy ambient medium and large-scale intracluster winds. Such winds are expected to bend the jets and distort the cocoon, potentially enabling backflows from opposite lobes to interact, mix, and most likely to reconnect conditions that may more closely reproduce the complex morphology observed in ESO~137$-$006. Incorporating cosmic-ray transport will provide a more complete description of how magnetised plasma structures evolve, dissipate  and survive within the intracluster environment, offering deeper insight into the physical origin of collimated synchrotron threads.

A complementary line of future work will assess the observational feasibility of detecting the filamentary structures identified in our GRMHD and GRRT models with next-generation interferometers. To this end, we will employ the \texttt{Karabo} pipeline\footnote{\url{https://i4ds.github.io/Karabo-Pipeline/index.html}}
 to create synthetic observations using our GRRT images as ground-truth sky models, sampled with the planned SKA-MID AA* configuration. The excellent sensitivity and comprehensive baseline distribution of AA*will allow us to evaluate whether the narrow magnetic filaments, shear-driven layers, and inward-moving, re-accreted flux tubes predicted by our simulations can be reliably recovered under realistic observational conditions. By comparing the reconstructed synthetic images directly with the original GRRT maps, we will quantify which aspects of the filament morphology remain robust after interferometric sampling, thermal noise, and image reconstruction. Ultimately, these synthetic-imaging studies will provide a crucial bridge between high-fidelity numerical models and forthcoming SKA-MID observations, offering a pathway to confirm or challenge the physical mechanisms responsible for the complex emission structures including magnetic filaments and collimated synchrotron threads revealed in current \textit{MeerKAT} observations and future SKA observations.

\bibliographystyle{abbrvnat-maxbibnames4}
\bibliography{chapter} 

\end{document}